\documentclass[journal]{IEEEtran}
\ifCLASSINFOpdf
  \usepackage[pdftex]{graphicx}
\else
  \usepackage[dvips]{graphicx}
\fi
\pdfoutput=1

\usepackage{amsmath}
\begin{document}
%
\title{Real time data analysis with the ATLAS Trigger at the LHC in Run-2}

\author{Pierre-Hugues~Beauchemin\\
        (On behalf of the ATLAS Collaboration)
\thanks{Manuscript received June 22, 2018; revised March 29, 2019.}
\thanks{Tufts University, 574 Boston  Avenue, Medford, Massachusetts 02155, USA,
e-mail: hugo.beauchemin@tufts.edu.}
}


%


\maketitle

\begin{abstract}
The trigger selection capabilities of the ATLAS detector have been significantly enhanced for the Large Hadron Collider (LHC) Run-2 in order to cope with the higher event rates and with the large number of 
simultaneous interactions (pile-up) per proton-proton bunch crossing. A new hardware system, designed to analyse real time event-topologies at Level-1 came to full use in 2017. A 
hardware-based track reconstruction system, expected to be used real-time in Run-3, is designed to provide track information to the high-level software trigger at its full input rate. The 
high-level trigger selections are largely relying on offline-like reconstruction techniques, and in some cases multi-variate analysis methods. Despite the sudden change in LHC 
operations during the second half of 2017, which caused an increase in pile-up and therefore also in CPU usage of the trigger algorithms, the set of triggers (so called trigger menu) 
running online has undergone only minor modifications thanks to the robustness and redundancy of the trigger system and the use of a levelling luminosity scheme in agreement 
with LHC and other experiments. This paper gives a brief yet comprehensive review of the real-time performance of the ATLAS trigger system in 2017. Considerations will be 
presented on the most relevant parameters of the trigger (efficiency to collect signal and output data rate) as well as details on some aspects of the algorithms which are run real-time 
on the High Level Trigger CPU farm.\end{abstract}



%
\IEEEpeerreviewmaketitle

\section{Introduction}
\IEEEPARstart{O}{ne} of the main objectives of the Large Hadron Collider (LHC) physics program in Run-2 is to discover new phenomena beyond the Standard Model (SM) of particle physics. Following the tight constraints
set on many Beyond the Standard Model (BSM) scenarios by the Run-1 LHC analyses, a large spectrum of still viable BSM models require very exclusive final states in small regions
of the phase space. Another central piece of the LHC physics program is the precision measurement of electroweak and quantum chromodynamics (QCD) processes, as well as a complete mapping of the
various Higgs boson couplings and parameters. In order to meet these research objectives, data samples with very high statistics must be used for the searches and the measurements.
To maximize the amount of data collected by the detectors, the LHC is constantly increasing the luminosity delivered to the experiments. This constitutes a challenge for the trigger
and data acquisition (TDAQ) system of the LHC experiments because of the limited CPU and storage available.

During Run-1, the ATLAS trigger system operated efficiently primarily at center-of-mass energies of 7 
and 8 TeV and at instantaneous luminosities of up to $8\times10^{33}$ cm$^{-2}$s$^{-1}$\footnote{For a full description of the ATLAS detector, see~\cite{bib:atldetec}}. In Run-2, the center-of-mass energy increased to 13 TeV, enhancing the total proton-proton (pp) cross section by more than a factor of two, therefore increasing the trigger rate by more than 100\%. In addition, changes in the LHC beam parameters resulted in an increase of the instantaneous luminosity 
up to a factor 3, with up to 80 pp-interactions per bunch-crossing (in-time pile-up)  in 2017. Finally, a reduction of the bunch spacing from 50 ns to 25 ns 
added interactions from neighboring bunch-spacing (out-of-time pile-up). These changes in the LHC operation, designed to allow for the experiments to take data samples with larger statistics,
made the Run-1 trigger menu completely unsustainable. To preserve the physics program of the experiment, a significant upgrade of the ATLAS trigger system was needed for Run-2.

\section{Run-2 Improvements of the ATLAS Trigger System}
\label{sec:tdaq}

Improvements of the hardware, firmware and software parts of the trigger system must aim at a better rate control and processing time per event, higher reconstruction and identification 
efficiencies with respect to offline selections, and resolution effects closer to offline measurements. A detailed description of the upgrade of the trigger system is presented in~\cite{bib:atltriperf2015}.

\subsection{Level 1 Trigger Improvements}

From the hardware perspective, a fourth layer of resistive plate chambers (RPC) was added, before Run-1, to the muon spectrometer in order to recover acceptance lost at the first trigger level (L1) near 
detector feet and elevator shafts. These chambers were however only equipped with electronics during the long shutdown following Run-1. A net increase of 3.6\% in the muon L1 trigger 
efficiency resulted from this hardware addition while reducing the trigger rate by 60\%, thanks to its impact on the suppression of particles not originating from the interaction point.



Multiple changes have also been brought to the hardware and firmware L1 trigger system. A new Fast Tracking reconstruction system (FTK) has been developed~\cite{bib:ftk}, and will become 
fully operational in Run-3. The FTK system provides global inner detector (ID) track reconstruction at L1, using lookup tables in associative memory chips for pattern recognition. 
This FPGA-based track fitter performs a fast linear fit and the tracks are made available to the High-Level Trigger system (HLT). The FTK allows for the use of tracks at much higher event 
rates in the HLT than is affordable using CPU systems, improving, among others, the tau and the B-meson physics (B-physics) trigger performances. 

In order to refine the muon and calorimeter-based 
object kinematic calculations and to make more sophisticated event selections at L1 (e.g. invariant mass cuts, angular distance between jets, etc.), two FPGA-based processor modules 
(L1-Topo) have been added to the L1 trigger system, and became fully operational in 2017. Consequently, changes to the central trigger processor (CTP) were required (see~\cite{bib:atltriperf2015}
for details). The improvements made to the CTP and other L1 components allowed for a bunch-by-bunch pedestal subtraction that significantly reduced the rate of L1 jets and missing transverse 
energy ($E_T^{miss}$) triggers. It also linearized the L1 trigger rate as a function of the luminosity and the position of bunches in a train, and improved the bunch-crossing identification. 
Finally, the CTP upgrades allowed to double the number of L1 trigger signatures and bunch-group selections providing more sophisticated trigger chains for very exclusive 
event topologies. The improvements brought to the entire L1 trigger system allowed for a L1 accept rate of 100 kHz, which constitutes approximatively a 30\% increase with respect to the 
corresponding rate in Run-1. It also made it possible to keep a similar L1 trigger composition as in Run-1 despite the dramatic increase in the luminosity and pile-up.




\subsection{High Level Trigger Improvements}
\label{sec:hltimprove}

The entire High-Level Trigger (HLT) architecture has been changed after Run-1. The Level 2 and Event Filter farms have been merged to allow for more flexibility, to simplify the hardware and 
the software, and to remove rate limitations between fast and precision processing by using the resources more efficiently. 
To deal with the increase in the readout rate due to higher L1 accept rate, but to also increase the output rate of the TDAQ system, the Read-Out System (ROS) has also been upgraded. Thanks 
to these improvements, data have been stored at a rate of 1.1 kHz in Run-2, almost a factor of 3 increase with respect to Run-1~\cite{bib:atltriperf2015}. 

\begin{figure}[!t]
\centering
\includegraphics[width=3.2in]{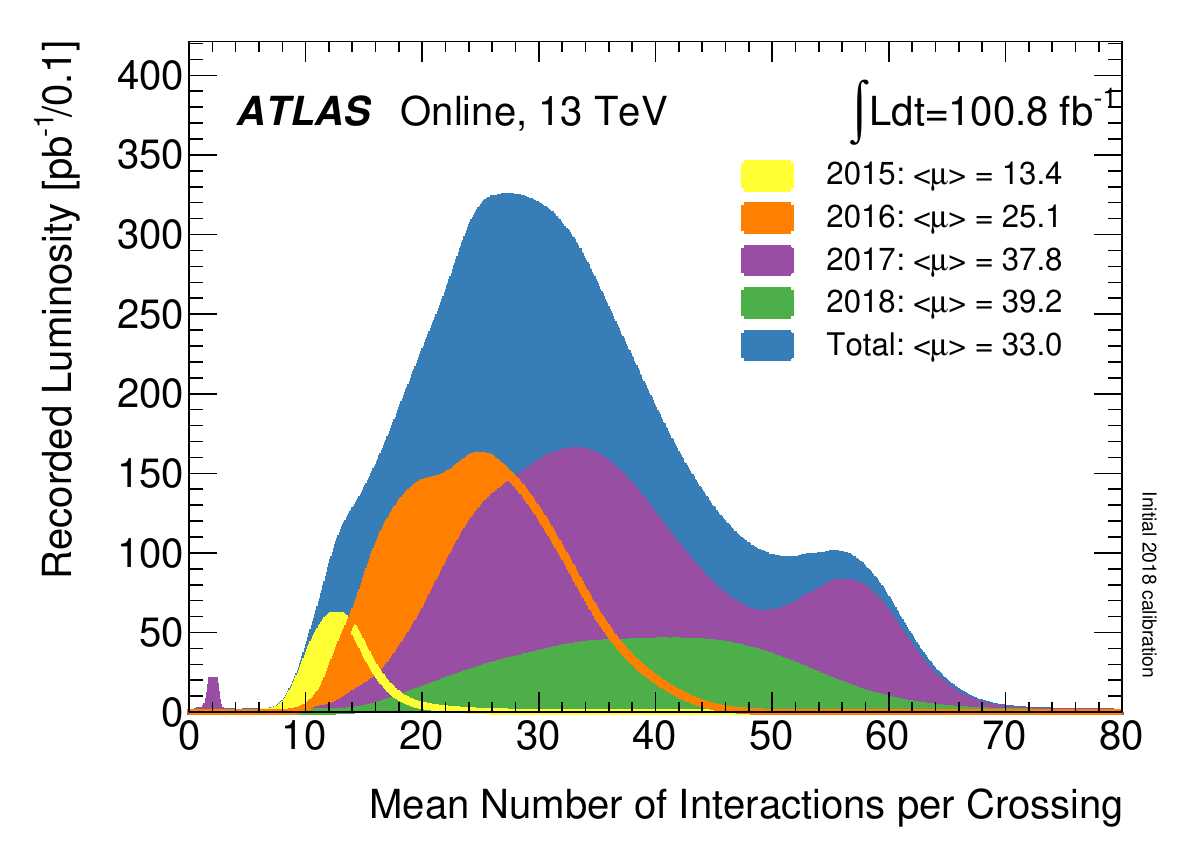}
\caption{Luminosity-weighted distribution of the mean number of interactions per bunch crossing for the 2015-2018 pp collision data at 13 TeV centre-of-mass energy. The mean number 
of interactions per bunch crossing corresponds to the mean of the poisson distribution of the number of interactions per crossing calculated for each bunch. It is calculated from the instantaneous 
per bunch luminosity as $\mu=L_{bunch}\times\sigma_{intel} / f_r$, where $L_{bunch}$ is the per bunch instantaneous luminosity, $\sigma_{inel}$ is the inelastic cross section which was taken to 
be 80 mb for 13 TeV collisions, and $f_r$ is the LHC revolution frequency~\cite{bib:trigpub}*.}
\label{fig:pileup}
\end{figure}

The output rate is however not the only limiting factor; the HLT processing time is also limited by the amount of CPU cores available at HLT. The time taken to process one event at the LHC is
determined by both the trigger menu and by the number of pile-up interactions which are continuously increasing with time, as can be seen in Fig.~\ref{fig:pileup}. At an instantaneous luminosity
of $5.2\times10^{33}$ cm$^{-2}$ s$^{-1}$ and an average pile-up of $<\mu> = 15$, the average HLT processing time is of 230 ms, which is well within the 2-3 seconds time available before time-out.
However, as was reported in~\cite{bib:atltriperf2015}, the average processing time increases with luminosity and pile-up, and the distribution of HLT processing time has a tail that goes well above 
the time-out threshold. Part of these long-processing events can be recovered thanks to the data stream procedure (debug stream), but there is an imperative need for HLT algorithms to cleverly deal 
with pile-up to avoid a significant decrease in the triggering performance.


\begin{figure}[!t]
\centering
\includegraphics[width=3.2in]{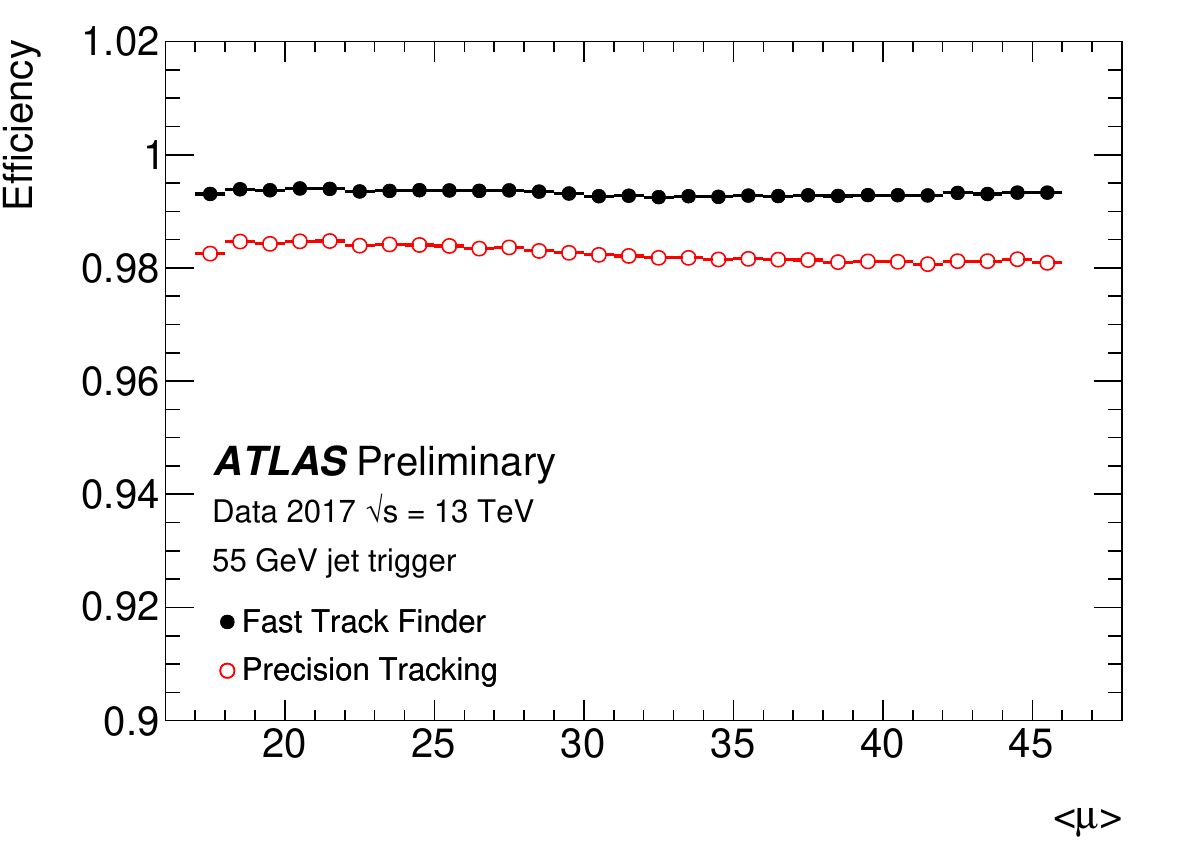}
\caption{The track finding efficiency of the Inner Detector (ID) trigger for tracks with $p_T > 1$ GeV within jets shown as a function of
the multiplicity 
of the mean number of pileup interactions in the event. For the jet and Bjet triggers the reconstruction in the ID trigger runs in three stages~\cite{bib:trigpub}.*}
\label{fig:tracking}
\end{figure}

\begin{figure}[!t]
\centering
\includegraphics[width=3.2in]{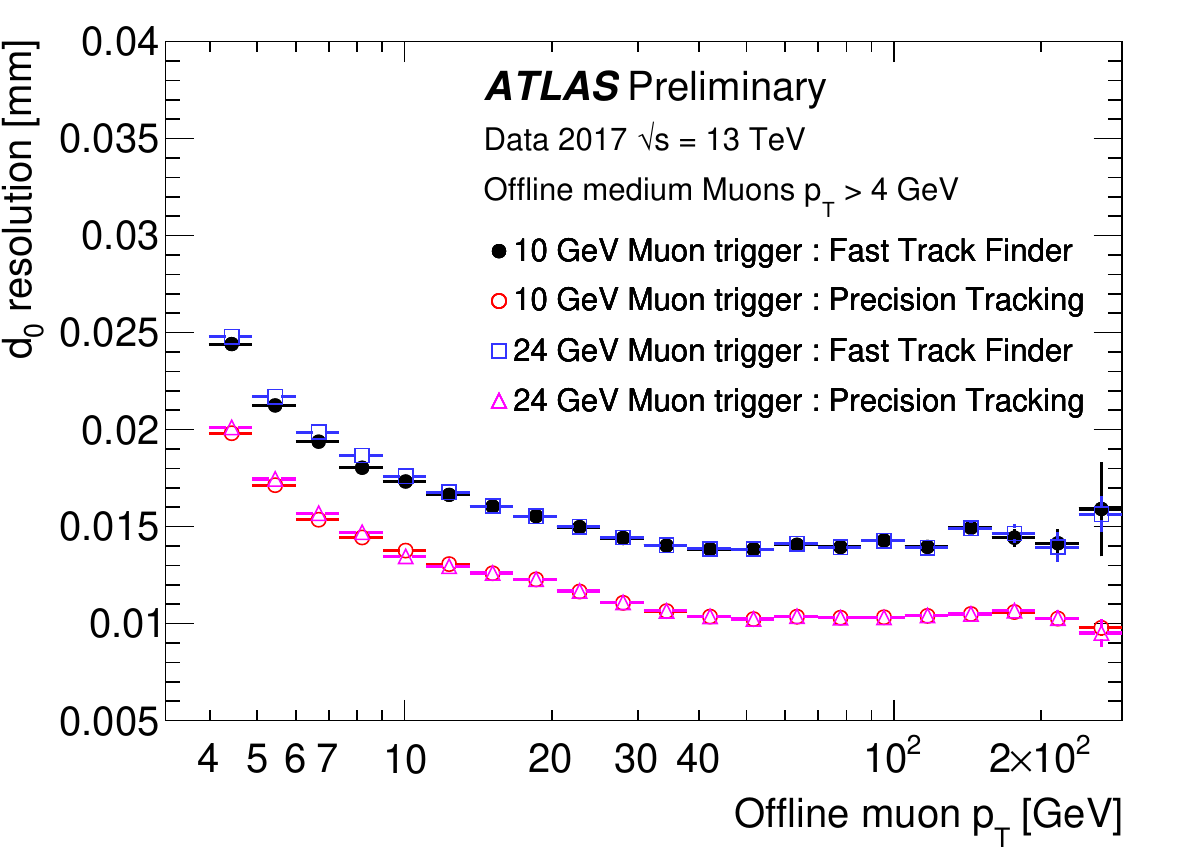}
\caption{
The trigger track transverse impact parameter resolution of the Inner Detector (ID) trigger for muons with $p_T > 4$ GeV from medium quality offline muon 
candidates, shown as a function of the offline muon $p_T$.  
This resolution is evaluated for both a 10 GeV and a 24 GeV muon triggers running in a mode where the trigger decision 
is made based on early muon candidates reconstructed from the Muon Spectrometer information only and so can contain candidates where the full offline reconstructed muons have a $p_T$ lower than 
the trigger threshold. The ID trigger first runs a Fast Track Finder stage followed by a detailed Precision Tracking stage to refine the track candidates identified in the first stage and improve their 
quality~\cite{bib:trigpub}*. }
\label{fig:muontracking}
\end{figure}

Many improvements have been brought to the online inner detector and muon spectrometer tracking~\cite{bib:atltriperf2015}. For example, to limit CPU usage, multiple stage track reconstruction was implemented, 
thanks to the redesign of the HLT architecture in Run-2. It allows, among other things, to use larger region of interest around L1 objects, to seed hadronic taus or b-quark jets (b-jets) reconstruct for example, before precision
tracking exploit aspects of offline tracking to improve resolution and reduce rate at no cost in efficiency, as can be seen in figures~\ref{fig:tracking} and~\ref{fig:muontracking}.

The signal output from the calorimeter readout is also processed to produce cells or clusters that are then used to reconstruct physics objects like electrons, photons, taus, jets and $E_T^{miss}$. The 
cells and the clusters are also used in the determination of the shower shape and isolation characteristics of these particles to enhance the purity of their identification. Two different clustering algorithms 
are used to reconstruct the clusters of energy deposition in the calorimeter: the sliding-window algorithm~\cite{bib:swalgo}, and the topo-clustering algorithm~\cite{bib:topoclus}. The first stage of their 
reconstruction consists in unpacking the data from the calorimeter. With the very high amount of pile-up events produced in Run-2, the possibility to reconstruct topoclusters for the full calorimeter on each 
event was compromised. However, with a new memory caching mechanism allowing for a very fast unpacking of the data, and with the development of offline-like clustering algorithms for HLT, the mean 
processing time for topo-clustering has been kept to 82 ms,
making topoclusters available to tau, jets and $E_T^{miss}$ algorithms on every events. An energy correction based on the bunch-crossing identification in a train and the average pile-up contribution 
for such bunch-crossing is applied to the topoclusters to reduce the out-of-time pile-up distortion of their 
energy measurement. As a consequence, the energy resolution of the topoclusters reconstructed online is comparable to what is achieved offline.

\begin{figure}[!t]
\centering
\includegraphics[width=3.6in]{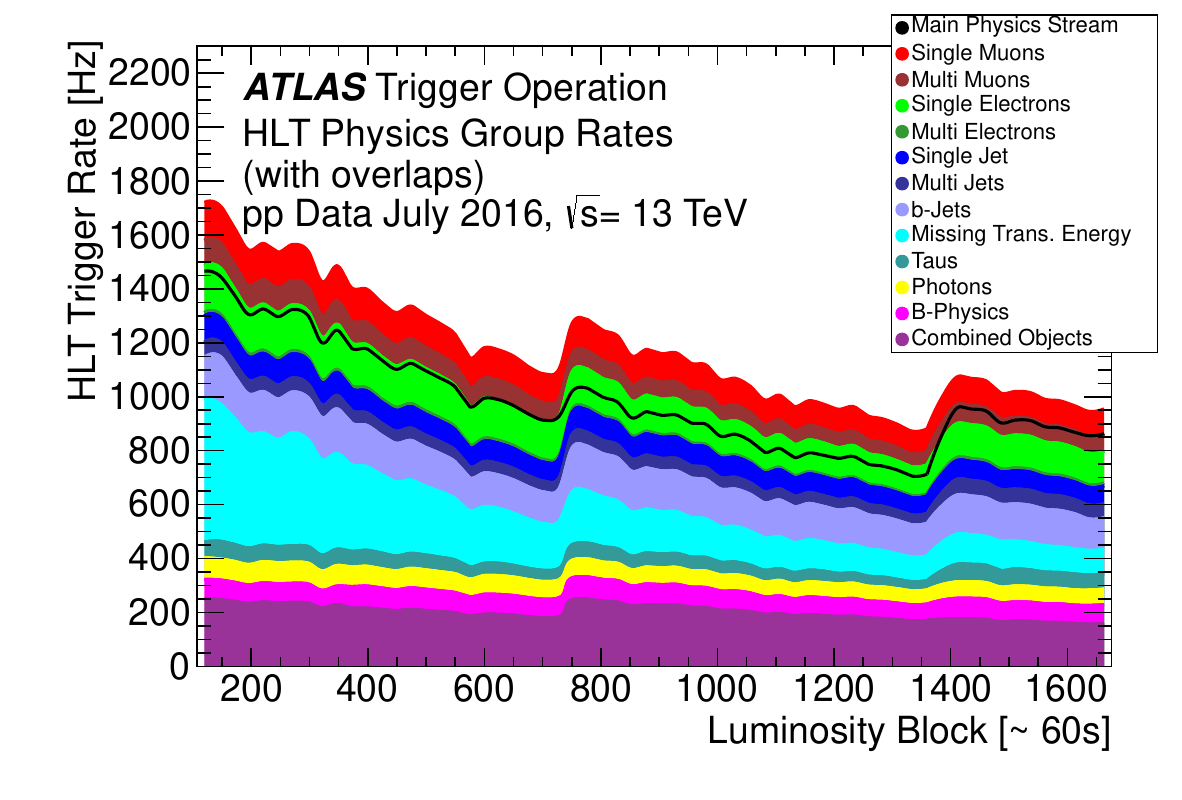}
\caption{HLT trigger rates grouped by trigger signature during an LHC fill in July 2016 with a peak luminosity of $1.02\times10^{34}$ cm$^{-2}$ s$^{-1}$~\cite{bib:trigpub}. Due to overlaps the sum of the individual groups 
is higher than the Main physics stream rate, which is shown as a black line. Multi-object triggers are included in the b-jets and tau groups. The B-physics triggers are mainly muon-based triggers. The 
combined group includes multiple triggers combining different trigger signatures such as electrons with muons, taus, jets or $E_T^{miss}$. Common features to all rates are their exponential decay with 
decreasing luminosity during an LHC fill. The rates periodically increase due to change of prescales to optimize the bandwidth usage, dips are due to deadtime, and spikes are caused by detector noise*.} 
\label{fig:menurate}
\end{figure}

Building on these various improvements, the trigger menu composition and the lowest transverse energy/momentum thresholds used by the different trigger objects for selecting events without random 
rejection (i.e. without prescale) has been designed to comply with the requirement of the LHC physics program. In 2015, it was even more inclusive than it was in Run-1. To maximize the output of the 
experiment relevant to the complete set of physics analyses to be carried during Run-2, the trigger menu is optimized for several luminosity ranges, changing even during a fill to use all available resources
while the instantaneous luminosity and  the average pile-up drop during a fill. An example of the bandwidth usage of the various triggering objects is presented in Fig.~\ref{fig:menurate}. 
Finally, the streaming strategy has been simplified: rather than using different data collection tags (streams) for events with muons (muon stream), electrons and photons (egamma stream), and jets, taus and $E_T^{miss}$ objects (JetTauETmiss stream), only one single Main Physics stream as been used to channel all the events to be
used in most physics analyses. This change reduced event duplication, thus reducing storage and CPU resources required for online reconstruction by roughly 10\%. In addition, a new streaming 
strategy, based on a partial event storage of only HLT reconstructed objects, sacrificing the ATLAS detector data needed for offline reconstruction, has been developed in Run-2. Such streams are used for 
calibration purposed, and to carry Trigger-Level Analyses (TLA)~\cite{bib:tla}. Such analyses are particularly useful for physics studies where the phase space probed is at kinematics lower than what is  
provided by the lowest unprescaled triggers. For example, more than one order of magnitude of dijet low $p_T$ events can be recovered by the Trigger-Level Analyses 
compared to what standard offline analyses can afford~\cite{bib:atltriperf2015}. This is another example of how creativity on real-time analysis serve the physics objective of the experiment. 

\section{Examples of Real Time Data Analyses}

Inner detector and muon spectrometer tracks, as well as calorimeter cells and clusters, are not directly used to select events at trigger level but are used as ingredients to reconstruct electrons, muons, taus, 
jets, b-jets, and $E_T^{miss}$ objects. In turn, these objects can be used to select multiple particle events, such as the triggers dedicated to B-physics for example\footnote{Note that some special triggers do 
not rely at all on the reconstruction of tracks or clusters, such as the random trigger, the minimum bias triggers, or the empty bunches triggers. These are not discussed here.}. The reconstruction and 
identification of these particles is critical for selecting as many events containing W bosons, Z bosons, H bosons and top-quarks as possible, on which most of the LHC precision measurements bare. The 
particles reconstructed at trigger level are also used to signal new phenomena. Inefficiencies in their reconstruction, or too high kinematic thresholds could compromise a BSM discovery or the precision 
of the SM parameters and cross sections to be obtained from these data. The task of the trigger is therefore to control rates in a way that keeps very high efficiency particle selections with thresholds as low
as possible. 

\begin{figure}[!t]
\centering
\includegraphics[width=3.2in]{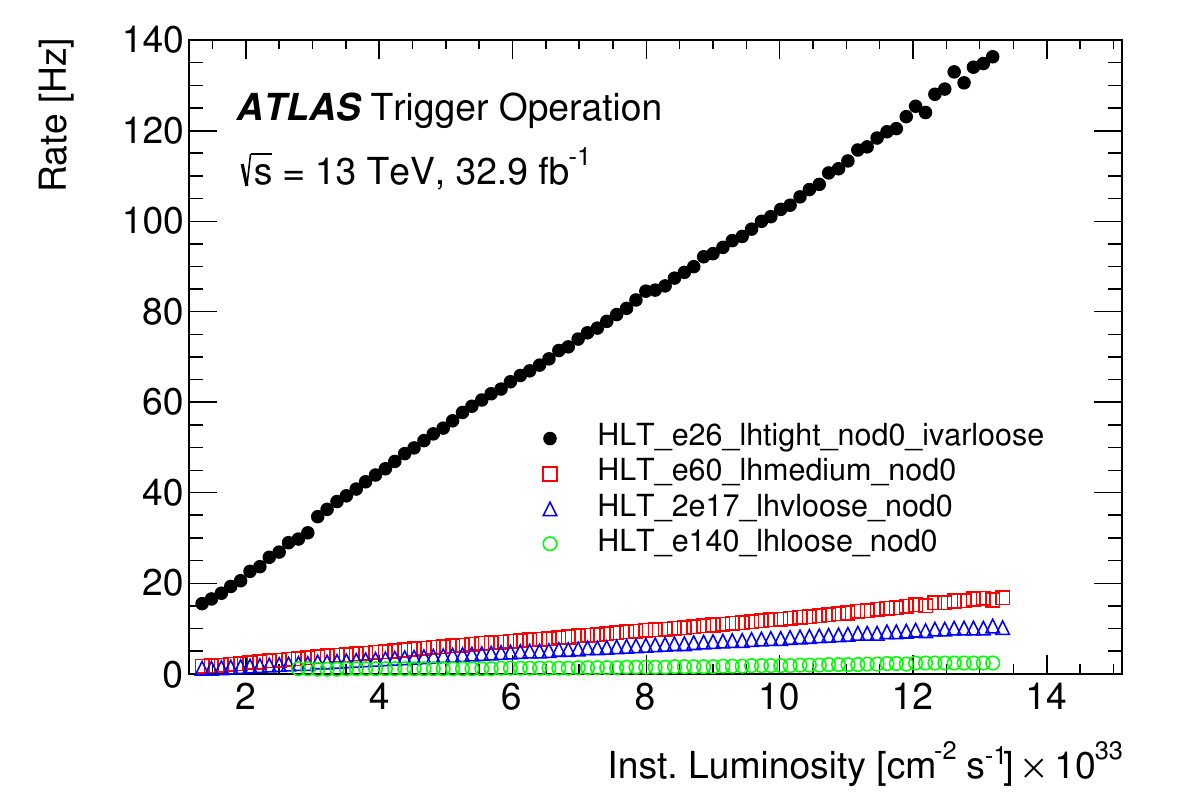}
\includegraphics[width=2.9in]{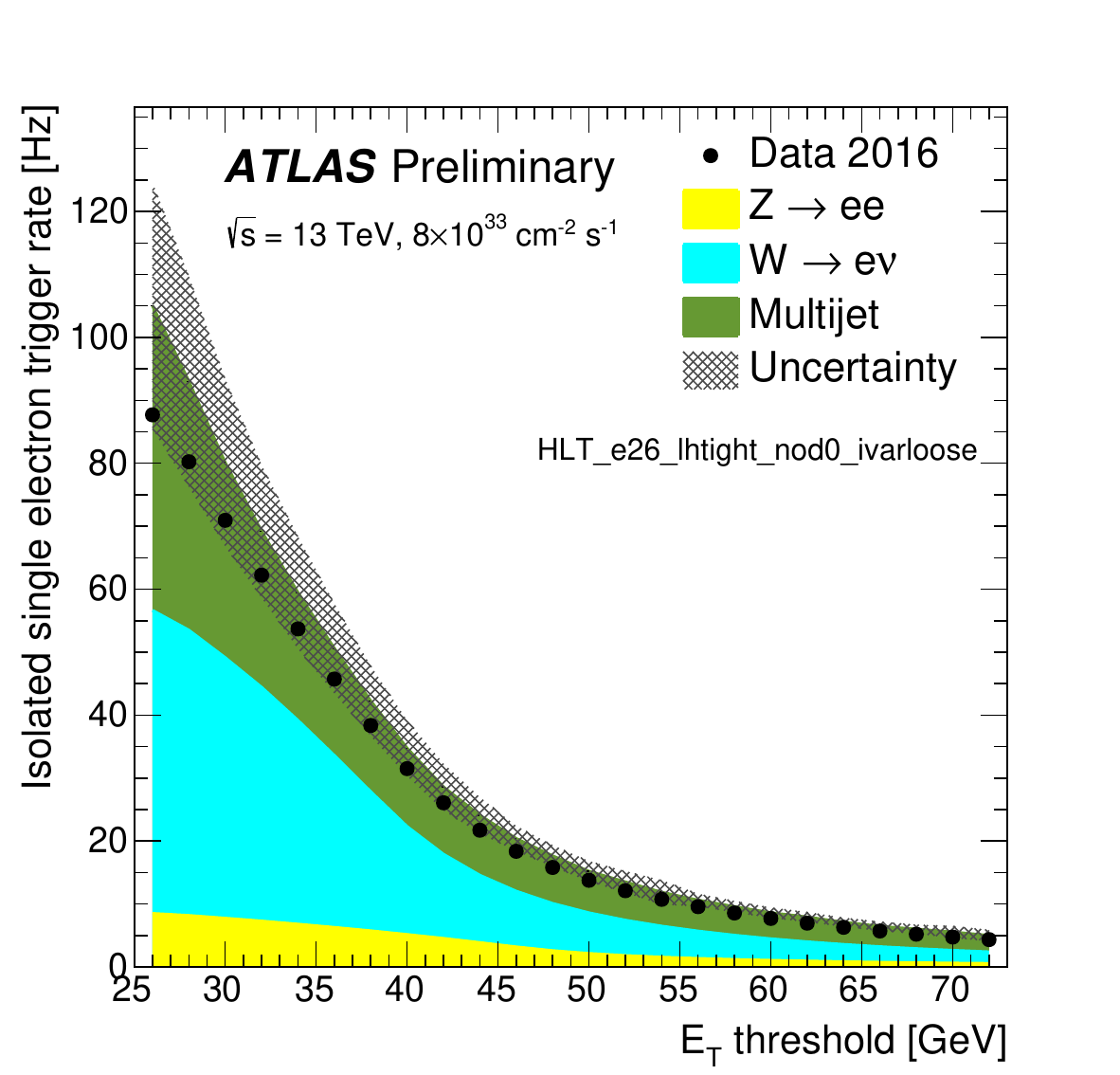}
\caption{{\bf Top}:  Output rates of the single-electron and di-electron primary triggers as a function of the un-calibrated instantaneous luminosity measured online during the 2016 proton-proton data taking at a 
center-of-mass energy of 13 TeV~\cite{bib:trigpub}*. {\bf Bottom}: Rate (in Hz) of the isolated single electron trigger as a function of the $E_T$ threshold at the high-level trigger (HLT) in the [26,72] GeV range, for the same likelihood-based 
tight identification and Level-1 selections. The rate is measured in a dataset collected at a constant instantaneous luminosity of $8\times10^{33}$ cm$^{-2}$ s$^{-1}$ at $\sqrt{s} = 13$ TeV, while the contribution from 
W, Z and multi-jet production is estimated with Monte Carlo. The dominant uncertainty on the multi-jet rate is evaluated with a data-driven technique~\cite{bib:trigpub}*.}
\label{fig:leptonrate}
\end{figure}

Because of the small cross section and the small fake rate for processes involving multiple reconstructed particles when one of them is an electron or a muon, the largest trigger bandwidth is attributed to the lowest 
unprescaled single lepton (electron, muon, and tau) triggers. This is illustrated in Fig.~\ref{fig:leptonrate} that gives an example of the single electron trigger rate compared to the dielectron trigger rate with even 
lower $p_T$ threshold on each electrons (top), as well as an estimate of the physics processes contributing to the lowest unprescaled single electron trigger used in 2016 (bottom). Similar patterns apply to muons. Jets are 
however more tricky. Because the LHC is a hadron collider, realm of the strong interaction, the production rate of dijet and multijet events is so large that the jet thresholds have to be very high to keep rates manageable. 
For example, the lowest unprescale single jet trigger has a threshold of 360 GeV, more than one order of magnitude larger than for the corresponding electron and muon triggers. Similar arguments apply to tau and 
$E_T^{miss}$. That was already the case in Run-1. However, in Run-2, these triggers had to also develop strategies to stay robust against pile-up in order to keep the thresholds relatively stable with respect to 
what they were in Run-1. Improvements in the different object reconstruction algorithms, building on the improvements of the overall trigger system presented in Sec.~\ref{sec:tdaq}, succeeded in meeting
the objective of keeping high efficiency at comparable kinematic thresholds as in Run-1 for all particles (electrons ($e$), muons ($\mu$), taus ($\tau$), etc.). Some of these successes are 
summarized below.  

\subsection{Electron trigger}
 
\begin{figure}[!t]
\centering
\includegraphics[width=3.2in]{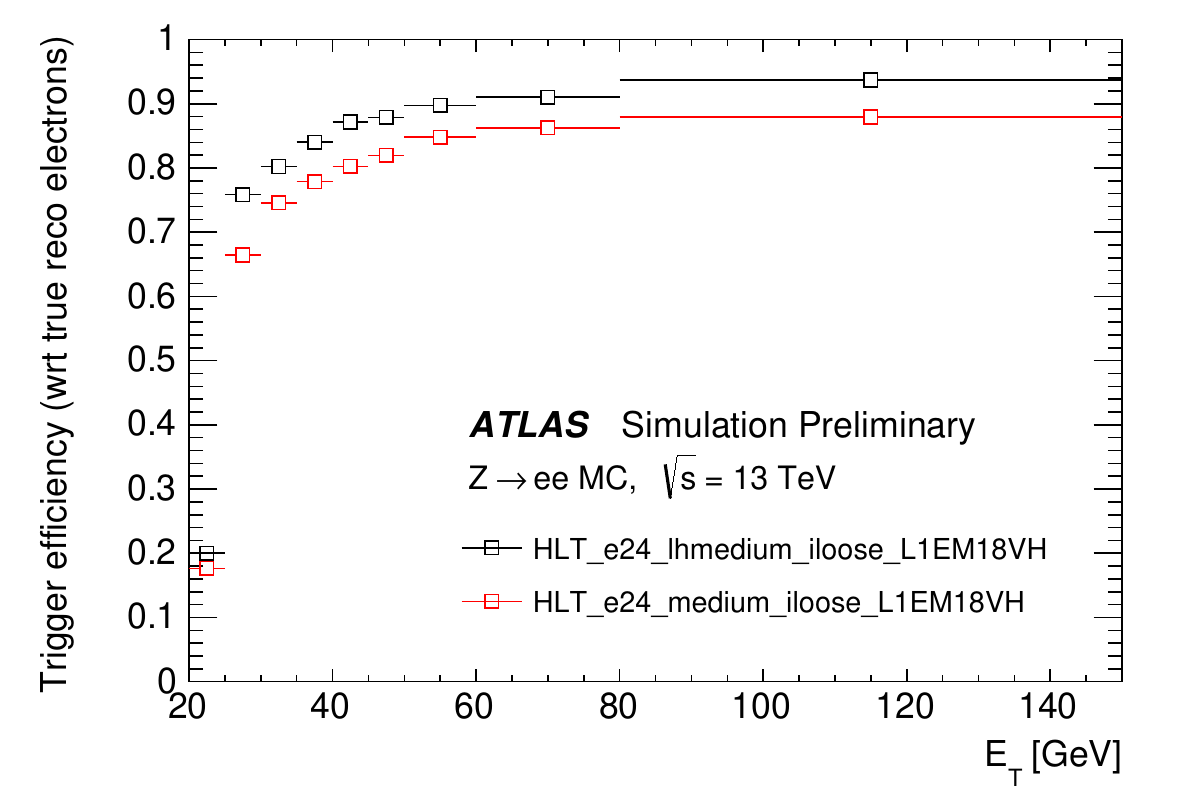}
\caption{Comparison of the likelihood-base and the cut-base HLT electron triggers efficiency as a function of the offline electron candidate's transverse energy $E_T$ with respect to true reconstructed electrons in 
$Z\rightarrow ee$ simulation. The HLT\_e24\_medium\_iloose\_L1EM18VH trigger is the Run-1 algorithm requiring an electron candidate with $E_T > 24$ GeV satisfying the cut-based medium identification, while 
HLT\_e24\_lhmedium\_iloose\_L1EM18VH corresponds to the Run-2 algorithm using the likelihood-based lhmedium electron identification. Both trigger chains also require the same track isolation selection and are
seeded by the same level-1 trigger (L1\_EM18VH)~\cite{bib:trigpub}*.}
\label{fig:lhvscut}
\end{figure}

\begin{figure}[!t]
\centering
\includegraphics[width=3.4in]{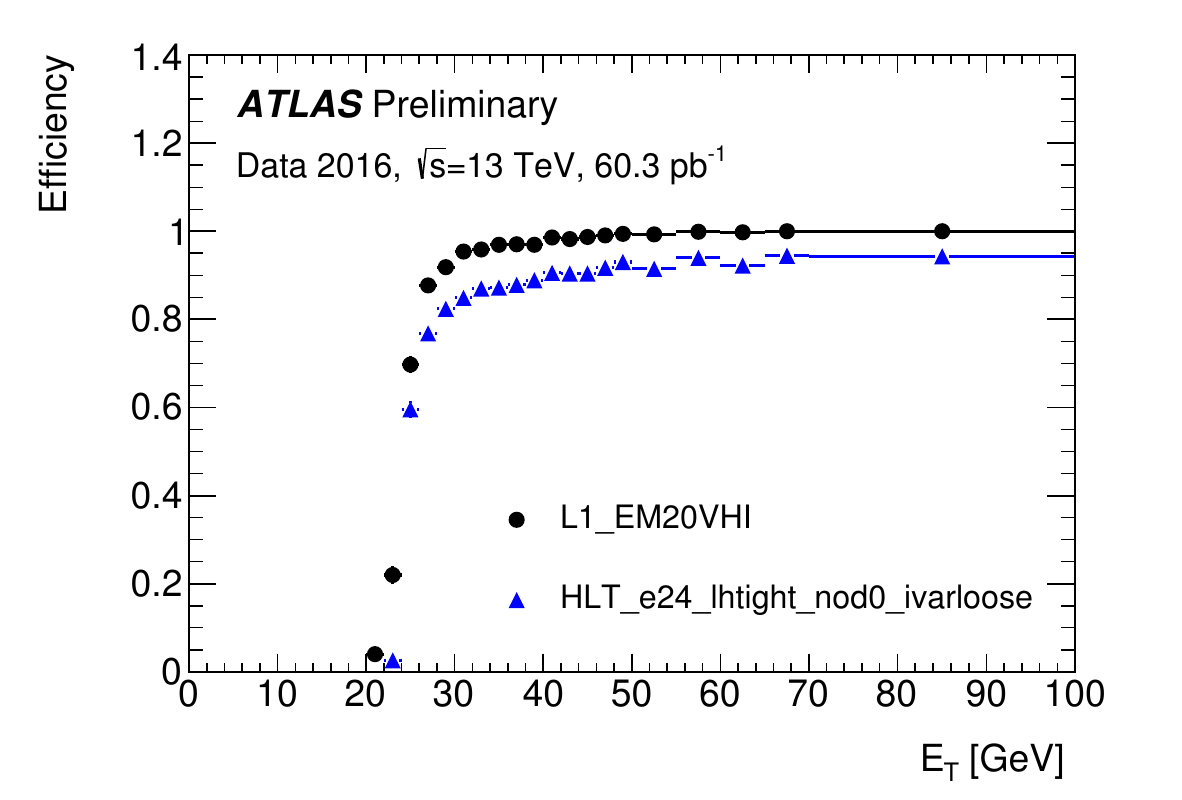}
\caption{Efficiency of the L1\_EM20VHI trigger (circles) as well as the combined L1\_EM20VHI and HLT\_e24\_lhtight\_nod0\_ivarloose trigger (blue triangles) as a function of the offline electron candidate's transverse 
energy ($E_T$).  A variable-size cone isolation criteria is applied ("ivarloose"). The HLT trigger requires an electron candidate with $E_T > 24$ GeV satisfying the likelihood-based tight identification. The offline reconstructed 
electron is required to pass a likelihood-based tight 
identification~\cite{bib:trigpub}*.}
\label{L1vsHLTegamma}
\end{figure}

\begin{figure}[!t]
\centering
\includegraphics[width=3.2in]{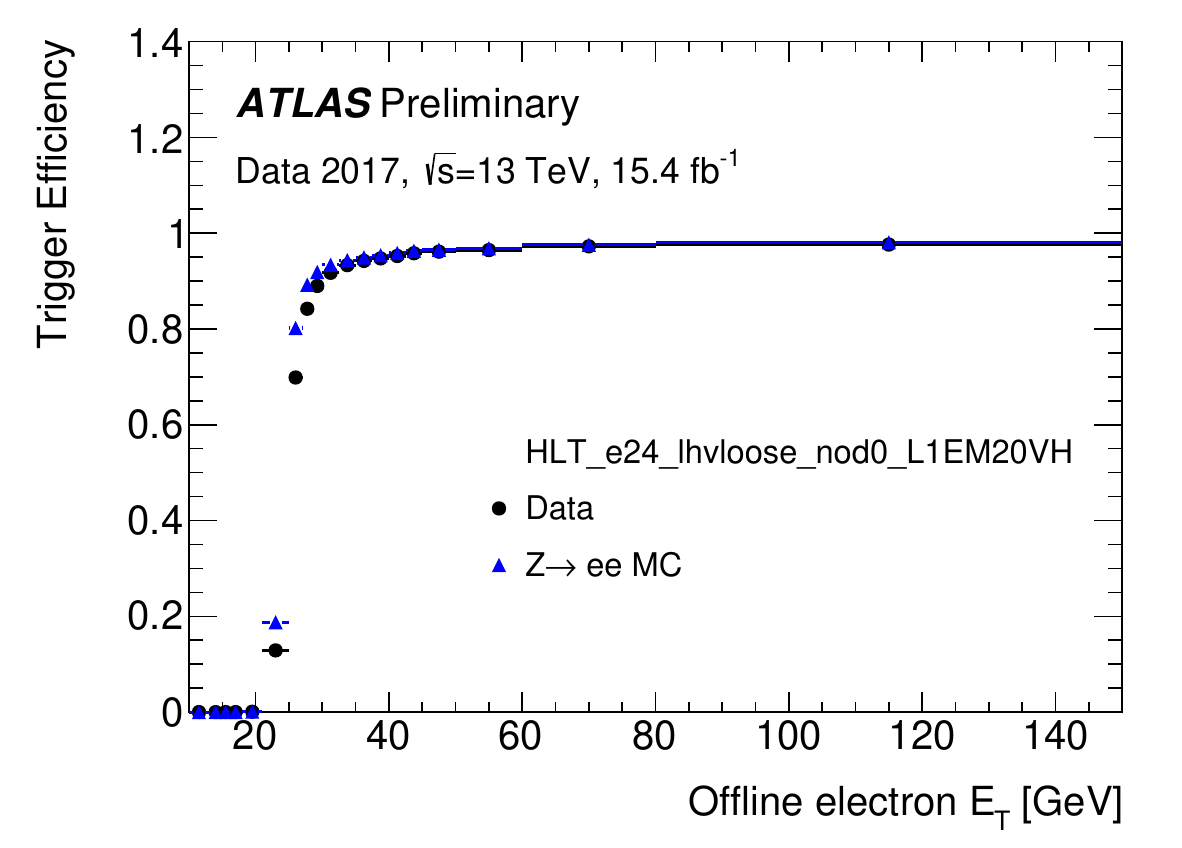}
\includegraphics[width=3.2in]{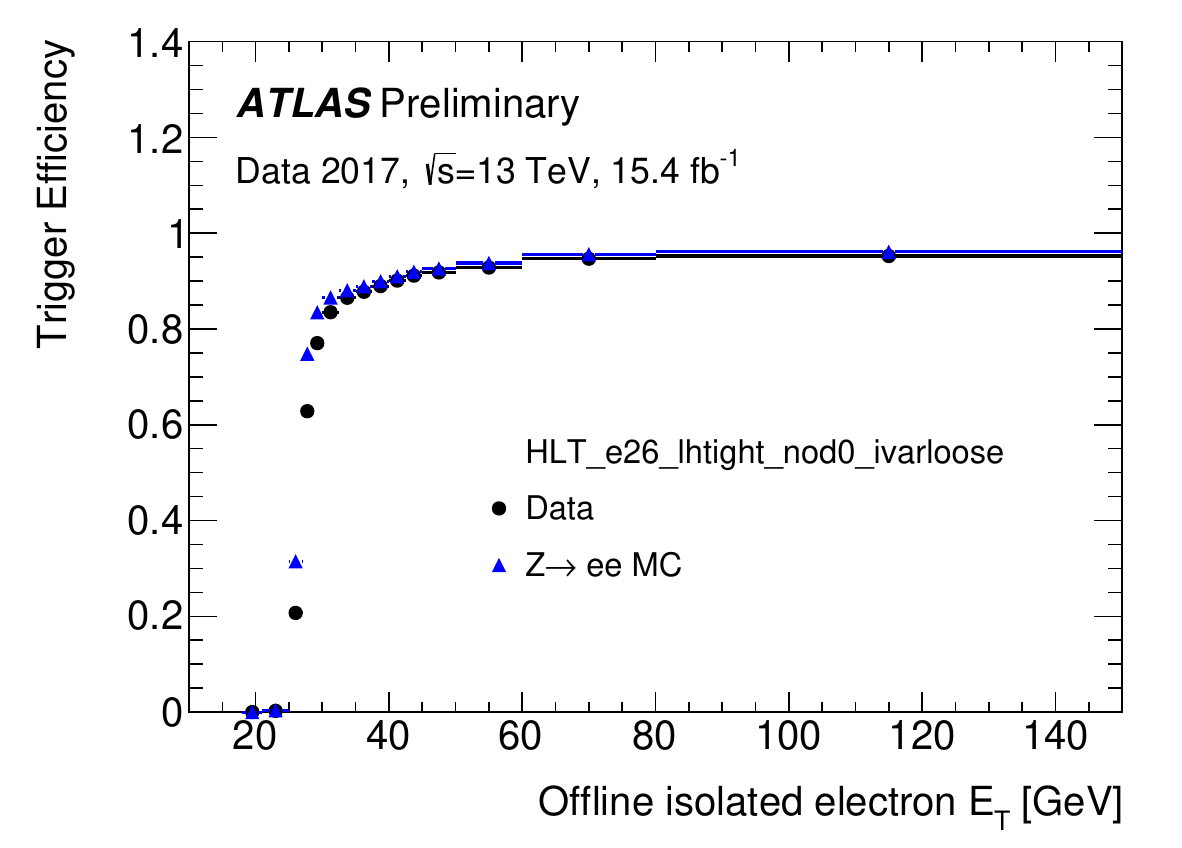}
\caption{Efficiency of the HLT likelihood-base electron trigger as a function of the offline electron candidate's transverse energy ($E_T$) as measured with the tab and probe method on a sample of 2017 ATLAS data as well
as on a $Z\rightarrow ee$ Monte Carlo sample~\cite{bib:trigpub}*. The data-to-MC agreement is very good. The efficiency is calculated for two different trigger chain: an electron candidate with $E_T > 24$ GeV satisfying the likelihood-based 
very loose identification and seeded by a 20 GeV electron L1 trigger ({\bf Top}); and an electron candidate with $E_T > 26$ GeV satisfying the likelihood-based tight identification and seeded by a 22 GeV electron L1 trigger ({\bf Bottom}). }
\label{2026egameff}
\end{figure}

 Like any other object, the objective of HLT electron reconstruction algorithms is to reject events as fast as possible, while identifying electrons and reconstructing their kinematics almost as efficiently and accurately as what 
 can be done offline. Because of the Run-2 improvements to the ATLAS TDAQ system, especially of the better CPU time management at the HLT, multivariate techniques are now being used online to a) calibrate the 
 energy of the clusters used to reconstruct electrons and photons; and b) implement the likelihood discriminant developed offline to identify electrons with a better purity vs efficiency figure of merit. Three working 
 points are used at HLT: loose, medium, and tight. The composition of the likelihood is the same as offline, with the exception of the momentum loss due to bremsstrahlung that is not accounted for in the online algorithm. 
 This approach has a better rejection for the same efficiency as the simple cut-base approach that was used in Run-1, or, conversely, a better efficiency for the same rate, as is demonstrated in Fig.~\ref{fig:lhvscut} with 
 simulation. Because of the high rejection rate of these electron identification algorithms, the lowest unprescaled $p_T$ threshold  at the HLT can be kept very close to the corresponding threshold at L1. This is impressive
 because the L1 accept rate is about 100 times larger than the HLT output. In Fig.~\ref{L1vsHLTegamma} we can see that the HLT efficiency turn-on curve is steeper than the L1 one. The cost for keeping such a low HLT 
 threshold is however that a 100\% efficiency is never reached by the HLT identification algorithm compared to offline. Comparing the top and bottom panel of Fig.~\ref{2026egameff} shows that the tighter is the identification
 working point, the larger is the signal efficiency lost. However, 
 increasing the $p_T$ threshold well beyond 30 GeV would compromise the precision measurements of many important physics parameters such as the W mass~\cite{bib:wmass}. It is therefore better to sacrifice a few percent efficiency for all $p_T$, to keep the bulk of the 
 electron $p_T$ distribution. These hard decisions have to be taken, while designing the trigger menu, by the whole ATLAS community, but the high performance of the trigger makes this decision easier. Note, as we 
 can see in Fig.~\ref{2026egameff}, that the data to Monte Carlo agreement is excellent, showing that we have a good understanding of the behavior of this trigger. 
 
 \subsection{Muon trigger}

\begin{figure}[!t]
\centering
\includegraphics[width=3.2in]{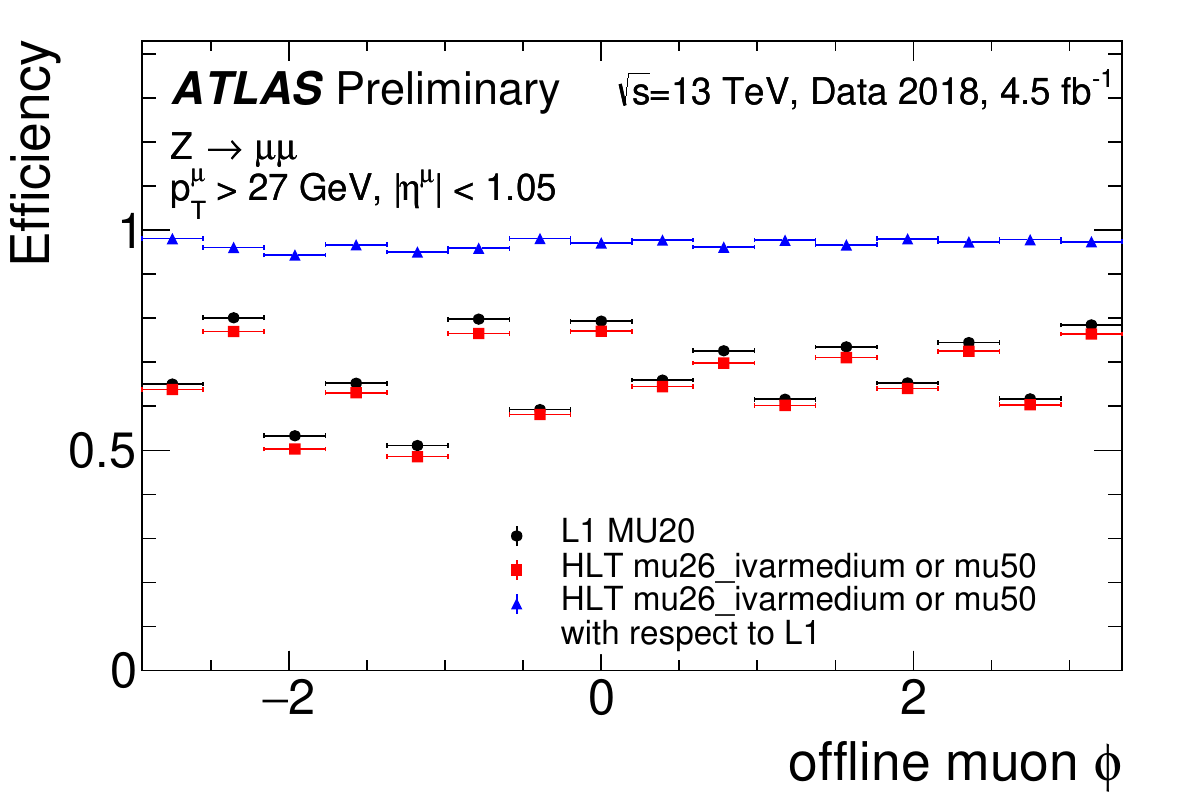}
\caption{Absolute efficiency of Level 1 (L1) MU20 trigger and absolute and relative efficiencies of the OR of mu26\_ivarmedium with mu50 High Level Triggers (HLT) plotted as a function of $\phi$ of offline muon candidates 
in the barrel detector region. The efficiency is computed with respect to offline isolated muon candidates which are reconstructed using standard ATLAS software and are required to pass "Medium" quality requirement. The 
selection is restricted to the plateau region with $p_T > 27$ GeV~\cite{bib:trigpub}*.}
\label{fig:muonphi}
\end{figure}

\begin{figure}[!t]
\centering
\includegraphics[width=3.2in]{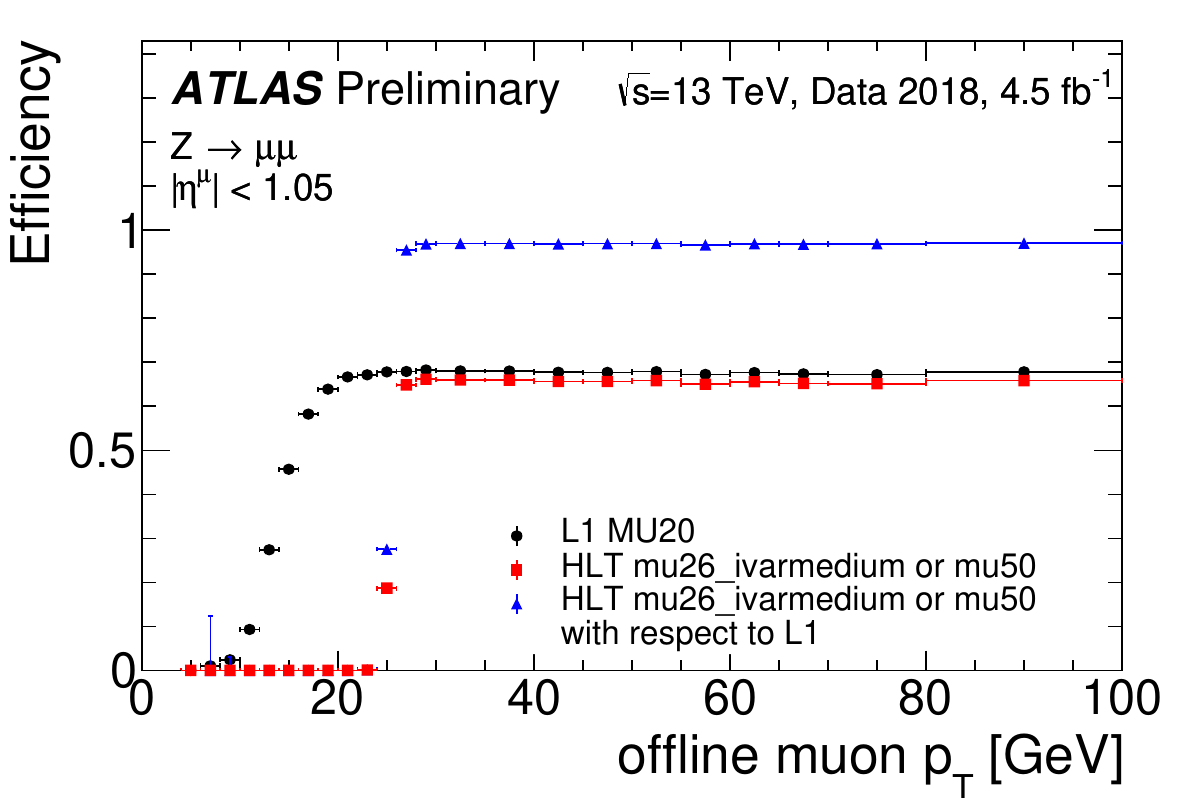}
\includegraphics[width=3.2in]{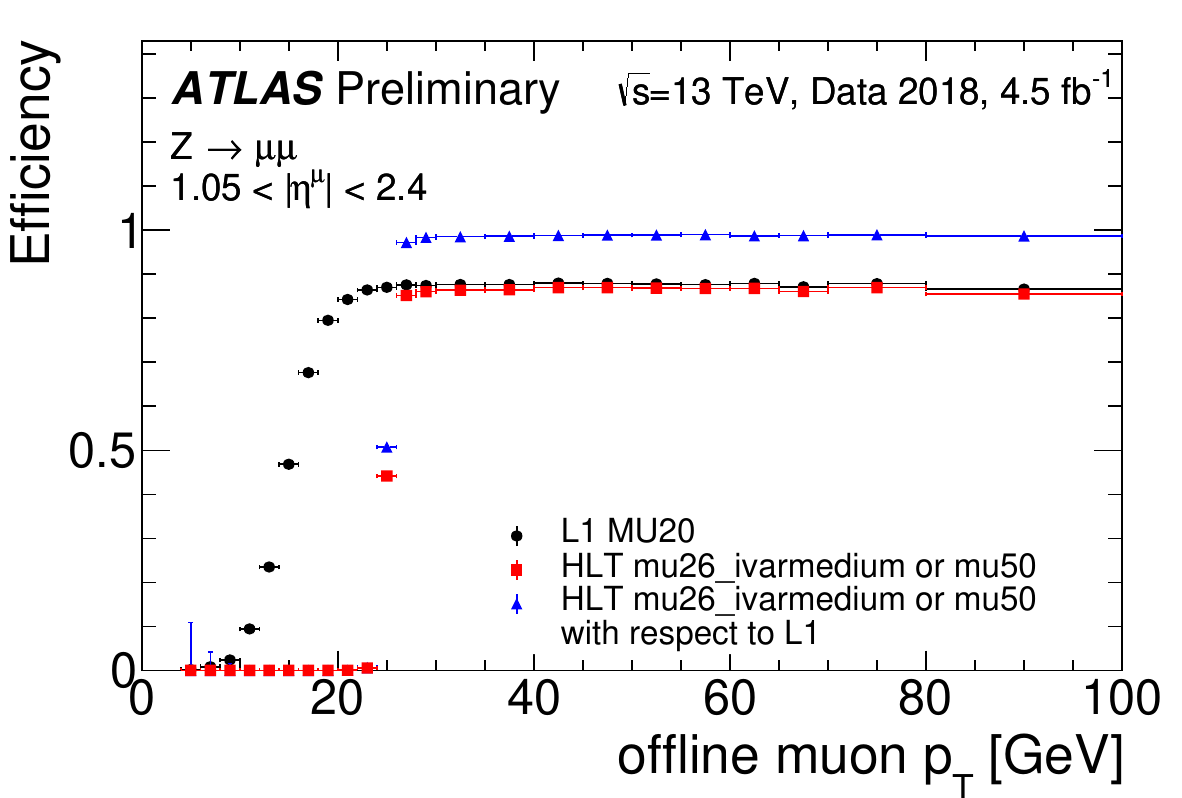}
\caption{Absolute efficiency of Level 1 (L1) MU20 trigger and absolute and relative efficiencies of the OR of mu26\_ivarmedium with mu50 High Level Triggers (HLT) plotted as a function of $p_T$ of offline muon candidates in the
barrel detector region ({\bf Top} ), and the endcap detector region ({\bf Bottom} )~\cite{bib:trigpub}*. The efficiency is computed exactly like described in the caption of Fig.~\ref{fig:muonphi}.}
\label{fig:muonpt}
\end{figure}

 For the muon triggers, the largest challenge is not so much the large background reduction, but the efficiency lost at L1 due to limited instrumentation coverage. As can be seen in Fig.~\ref{fig:muonphi}, there are significant
 variations of the L1 muon trigger efficiency as a function of the azimuth angle $\phi$ because of the limited RPC coverage for central rapidity ($|\eta|<1.05$) due to the detector feet, elevator shafts, and toroid magnets. We 
 can see that the HLT adds almost no inefficiency in selecting muons that can be reconstructed offline compared to the L1 trigger as HLT and offline both use the same detector signal. Fig.~\ref{fig:muonpt} presents the muon 
 trigger efficiency with respect to offline as a function of the transverse momentum of the offline muons. We can see on the top panel that the L1 inefficiency due to the lack of coverage of the RPC chambers amounts to about 30\%. 
 The problem is however about three times smaller for the region covered by the TGC detector, as can be seen on the bottom panel of Fig.~\ref{fig:muonpt}. In both cases we can see that the HLT-only muon trigger algorithm is 
 performing very similarly than the offline muon reconstruction and identification algorithm: the HLT turn-on curve with respect to L1 is very close to be a step function. Note that despite this limit of acceptance, the trigger
 and offline reconstruction algorithm are very precise and the background usually well understood, such that measurements in the muon channel are often the most precise. 
 
\subsection{Jet trigger}

\begin{figure}[!t]
\centering
\includegraphics[width=3.2in]{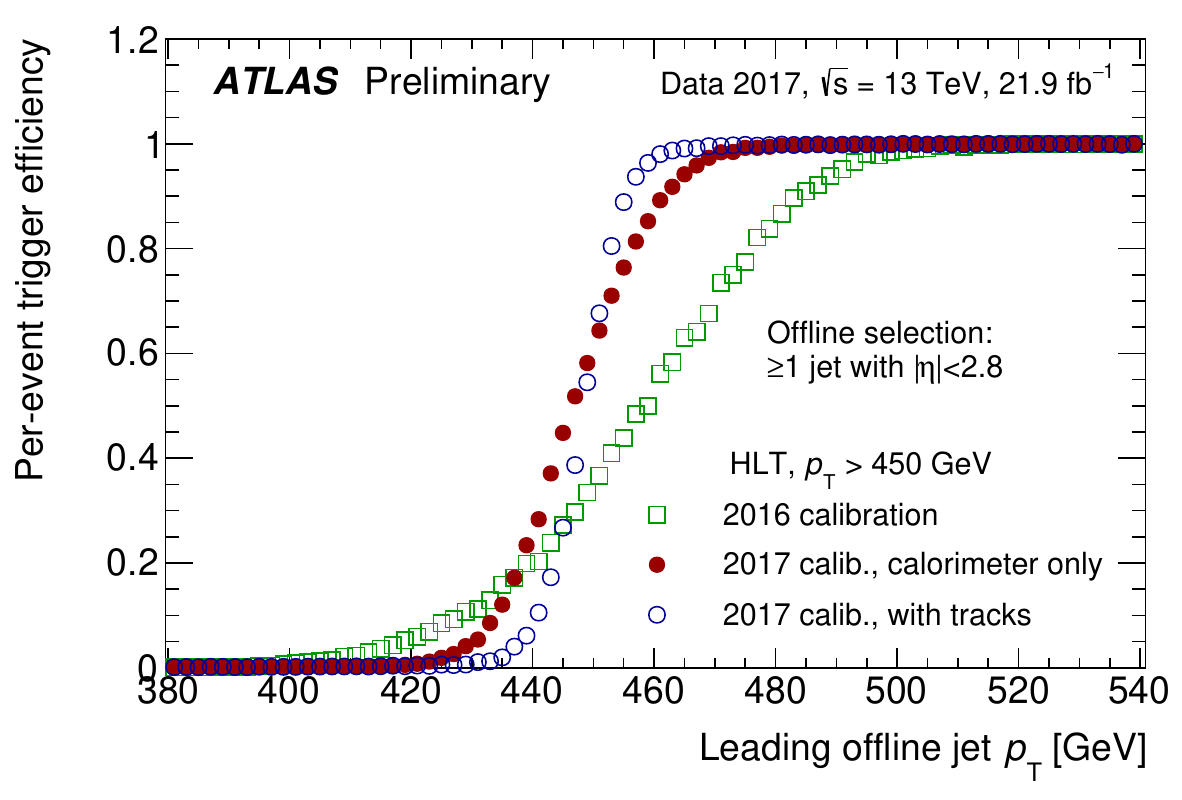}
\caption{Efficiencies are shown for a single-jet trigger with three different calibrations~\cite{bib:jetcalib} applied to jets in the ATLAS high-level trigger (HLT)~\cite{bib:trigpub}*.}
\label{fig:jeteff}
\end{figure}

\begin{figure}[!t]
\centering
\includegraphics[width=3.4in]{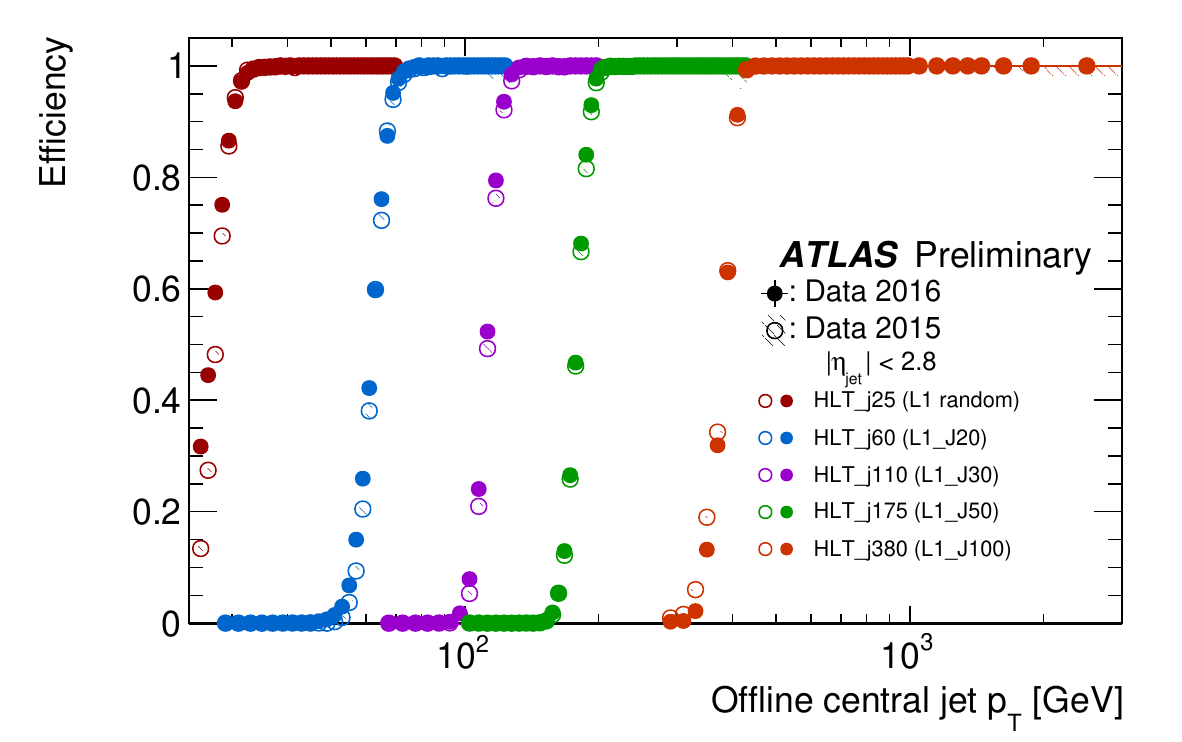}
\includegraphics[width=3.4in]{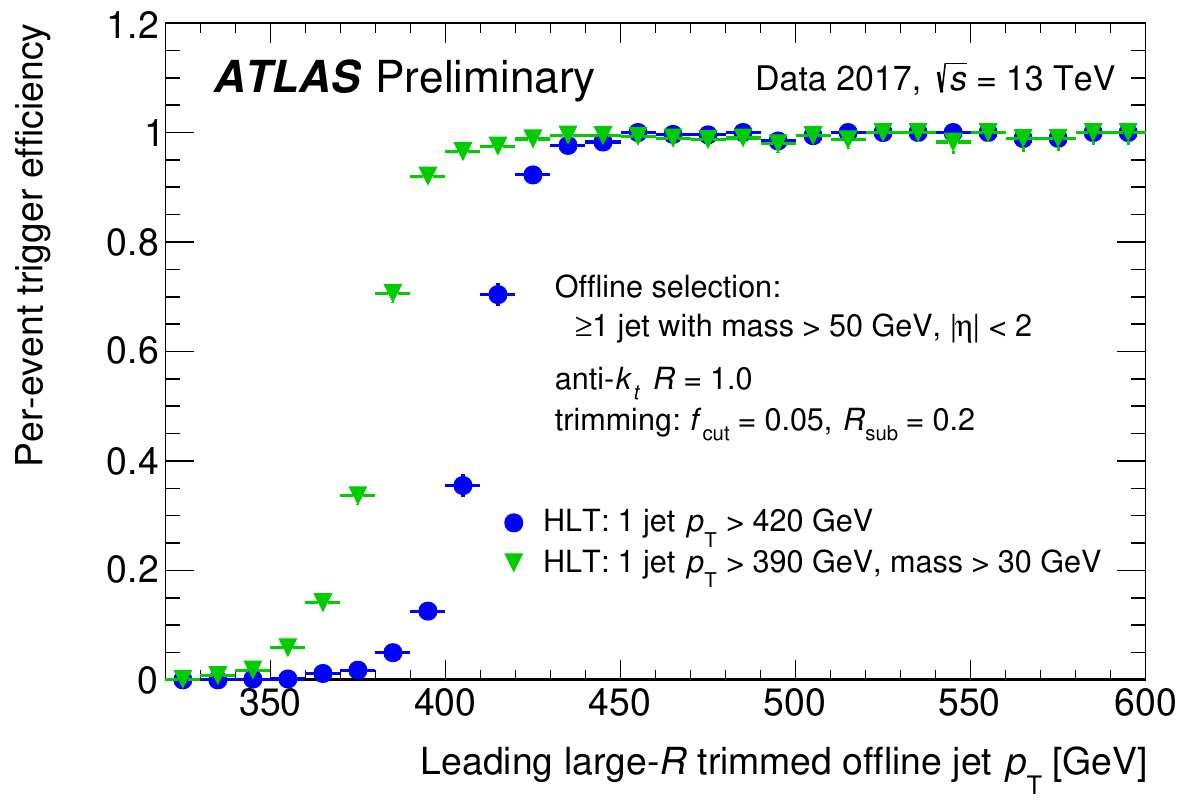}
\caption{{\bf Top}: Efficiencies for HLT single-jet triggers as a function of leading offline jet $p_{T}$. Triggers denoted HLT\_jX accept an event if a jet is reconstructed at HLT with $E_T > X$ GeV*.
The unprescaled trigger with the lowest threshold requires a jet with $E_\mathrm{T} > 380$ GeV~\cite{bib:trigpub}. {\bf Bottom}: Efficiencies for HLT large-R single-jet triggers as a function of the leading offline 
trimmed~\cite{bib:trim} jet $p_\mathrm{T}$. Blue circles represent a trimmed large-R jet trigger with a $p_\mathrm{T}$ threshold of 420 GeV. Adding an additional 30 GeV cut on the jet mass of the selected trimmed trigger jet is shown in green 
triangles. The mass cut significantly suppresses the QCD di-jet background, allowing a lower $p_\mathrm{T}$ threshold of 390 GeV, while retaining nearly all signal-like jets with a mass of above 50 GeV*.}
\label{fig:bootstrapjets}
\end{figure}

Beside the rate difficulties discussed above, the main challenges for jet triggers at the HLT are to preform an accurate calibration of the jets and to efficiently trigger on them in a high pile-up environment. As presented in 
Sec.~\ref{sec:hltimprove}, topoclusters, very similar to the offline ones, are used as input to the HLT jet algorithm. Jets are then calibrated in a two-step procedure similar to that adopted for offline analyses: first, pile-up contribution
is subtracted on an event-by-event basis using the calculated area of each jet and the measured energy density in the central part of the calorimeter; second, the response of the calorimeter is corrected using a series of $p_T$-
and $\eta$- dependent calibration factors derived from simulation. The calibration strategy is continually improving as can be seen in Fig.~\ref{fig:jeteff}. Starting in 2017, the calibration also uses track information. The sharp HLT 
efficiency turn-on curves presented in this figure prove that there is a good agreement between the HLT and the offline jet energy measurements. Note that on the contrary to electron and muon efficiency measurements, a bootstrap 
method~\cite{bib:jettrig} is used to obtain the jet trigger efficiency as is illustrated on the top panel of Fig.~\ref{fig:bootstrapjets}. Many physics analyses focus on events with heavily boosted massive particles decaying to multiple jets that are 
collimated. To avoid large efficiency lost due to jet reconstruction algorithm not adapted to this kind of event topologies, the jet reconstruction algorithm is fast enough to be run twice on an event in order to produce large size 
(large-R) jets from the output of the standard jet algorithm. Special jet trigger elements are then added to the menu to efficiently select such large-jet events. The performance of this jet algorithm is illustrated on the bottom panel of Fig.~\ref{fig:bootstrapjets}.

 \subsection{Tau trigger}

\begin{figure}[!t]
\centering
\includegraphics[width=3.3in]{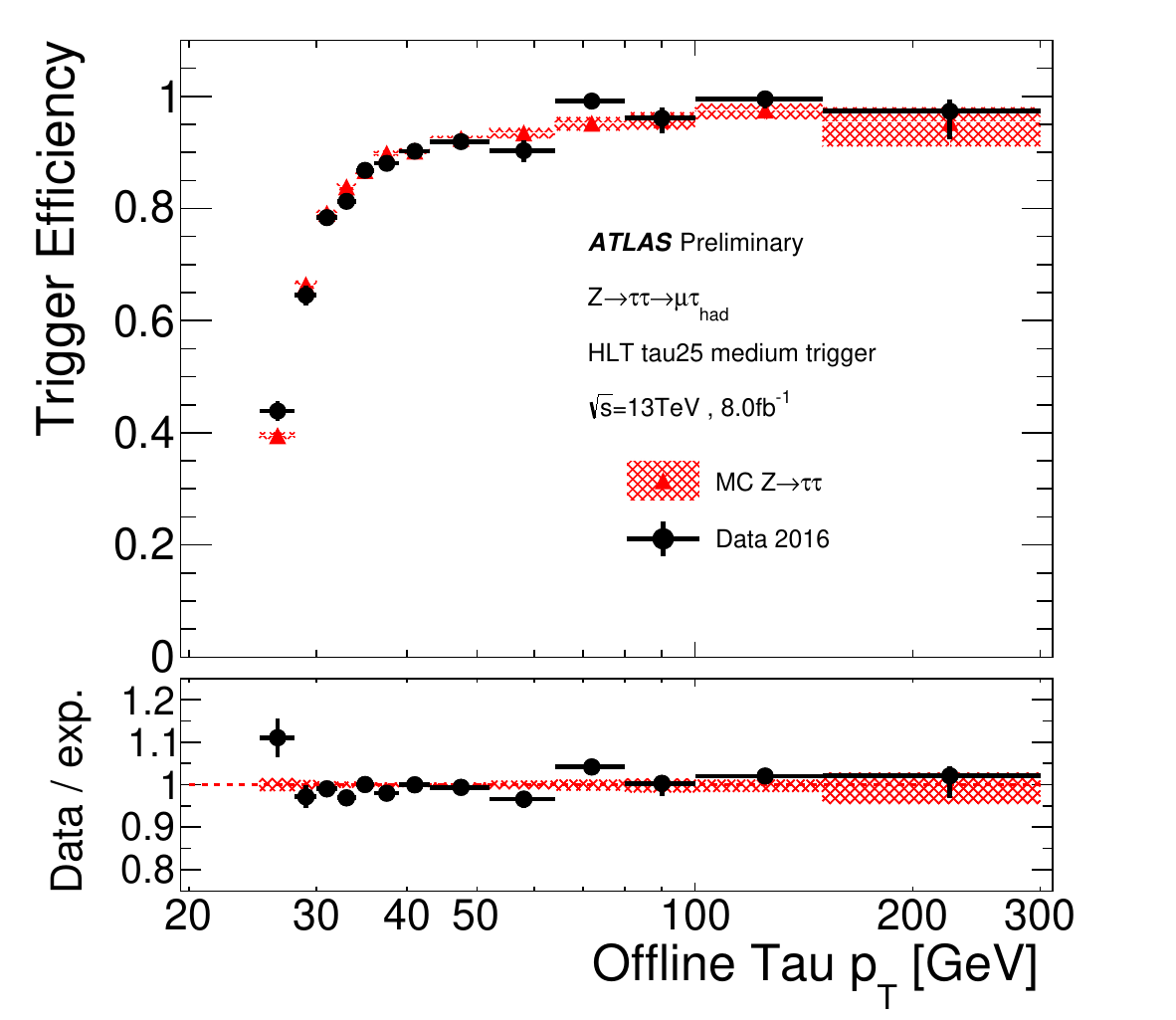}
\includegraphics[width=3.3in]{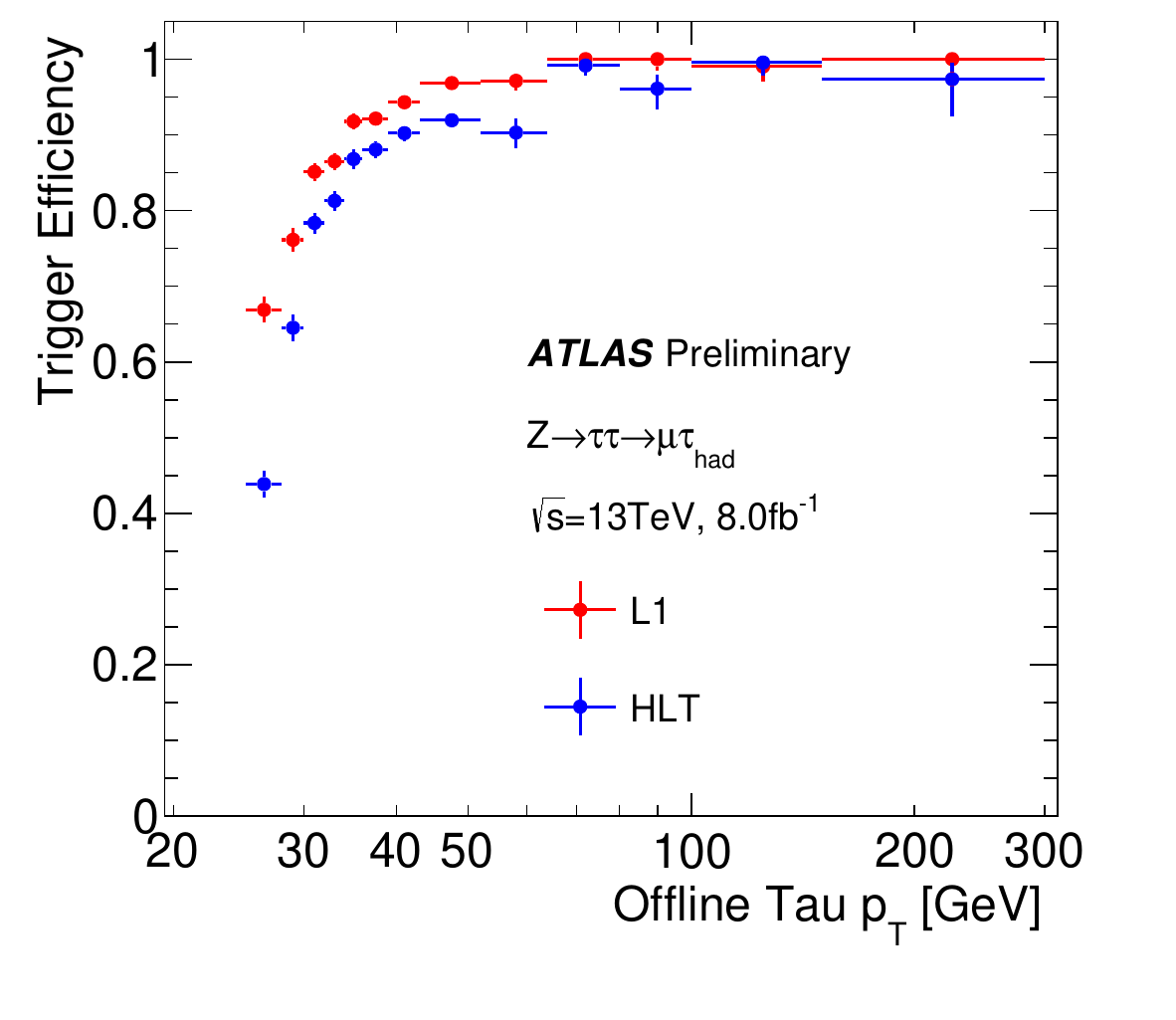}
\caption{{\bf Top}: Tau trigger efficiency measured in data and compared to simulation, with respect to offline reconstructed tau candidate with one or three tracks and passing the offline medium identification criteria, as function 
of the offline transverse momentum. The trigger efficiency is measured in a tag and probe analysis with $Z\rightarrow\tau\tau\rightarrow\mu\tau_{had}$ event from the 2016 dataset in 13TeV collision (8.0 fb$^{-1}$)*. 
{\bf Bottom} Comparison of this HLT tau trigger efficiency with the L1 tau trigger efficiency~\cite{bib:trigpub}*.}
\label{fig:taueff}
\end{figure}

While data samples enriched in leptonically decaying tau particles are selected by electron and muon triggers, hadronically decaying taus require a dedicated trigger. These are in essence narrow jets. Keeping the tau trigger rates under
control for $p_T$ thresholds low enough for the physics of interest is particularly challenging. To meet this objective, a 3-step reconstruction algorithm is deployed at the HLT. In the first step, narrow calorimeter energy deposits
are identified from the reconstructed topoclusters found in a cone of size $\Delta R = 0.2$ around the L1 object used to seed the HLT. In a second step, tau candidates are selected if there is a small number of reconstructed 
tracks pointing to the tau cluster, with the leading track central to it. Finally, a collection of variables built from the topoclusters and the tracks obtained by a precision tracking algorithm is used in a Boosted Decision Tree (BDT) 
multivariate algorithm to produce a score with which the final tau identification is done. To maximize the correlation between online and offline identifications, the BDT is trained using offline inputs. To mitigate pile-up effects, all 
variables used in the BDT are corrected according to the expected average interaction per bunch-crossing. 
Measurements of the tau trigger efficiency as a function of the offline tau $p_T$ have been obtained using the tag-and-probe technique 
on high purity samples. Results are presented in Fig.~\ref{fig:taueff}. As can be seen on this figure, the tau trigger efficiency is well-modeled by the Monte Carlo (top panel), and the HLT is only adding a marginal extra
source of inefficiency compare to L1 (bottom panel). 


 \subsection{$E_T^{miss}$ trigger}

\begin{figure}[!t]
\centering
\includegraphics[width=3.0in]{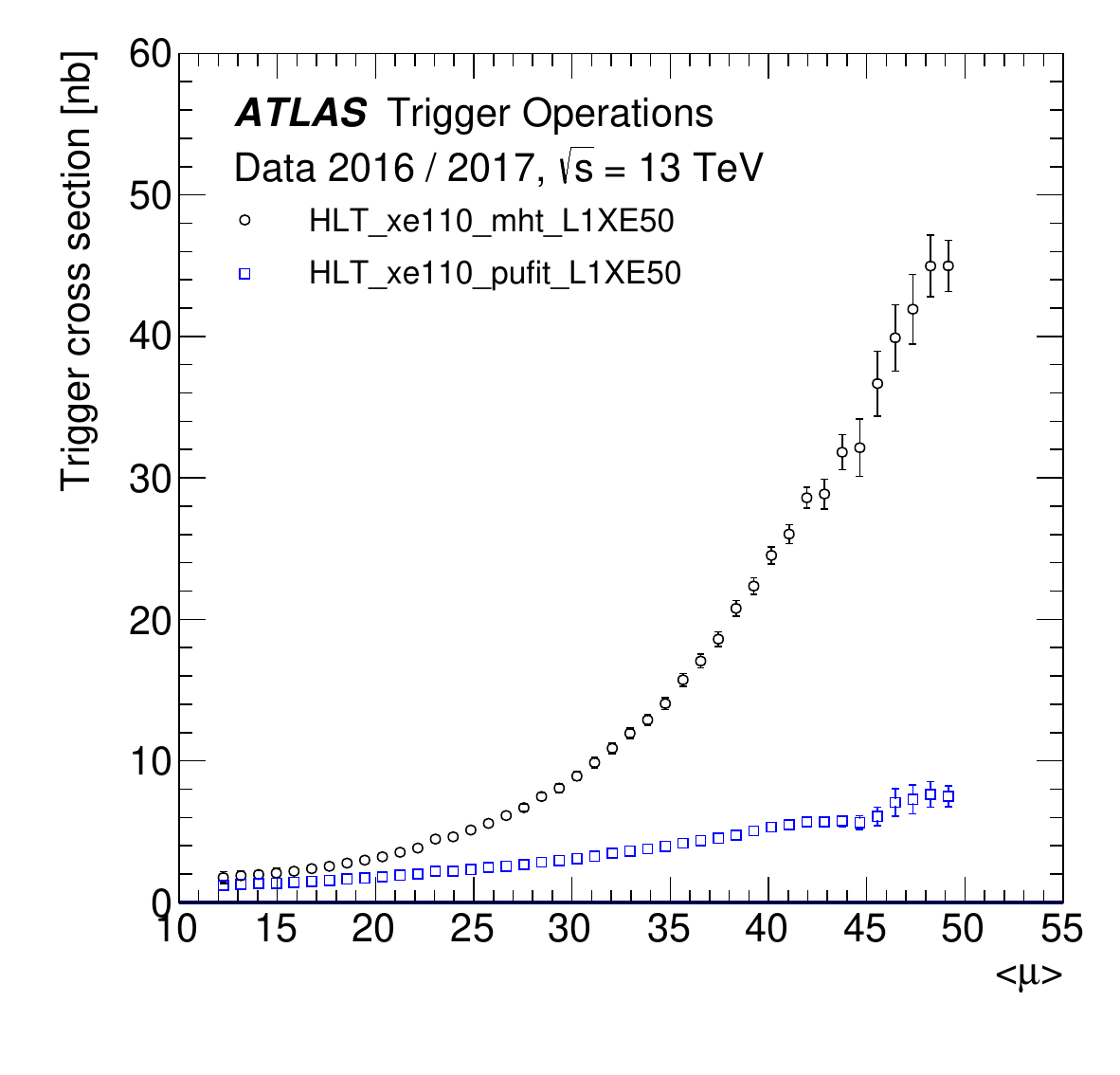}
\caption{The trigger cross-section as measured by using online rate and luminosity is compared for the main trigger $E_T^{miss}$ reconstruction algorithms used in 2016 ("mht") and 2017 ("pufit") as a function of the mean 
number of simultaneous interactions per proton-proton bunch crossing averaged over all bunches circulating in the LHC~\cite{bib:trigpub}*. }
\label{fig:pufit}
\end{figure}

\begin{figure}[!t]
\centering
\includegraphics[width=3.6in]{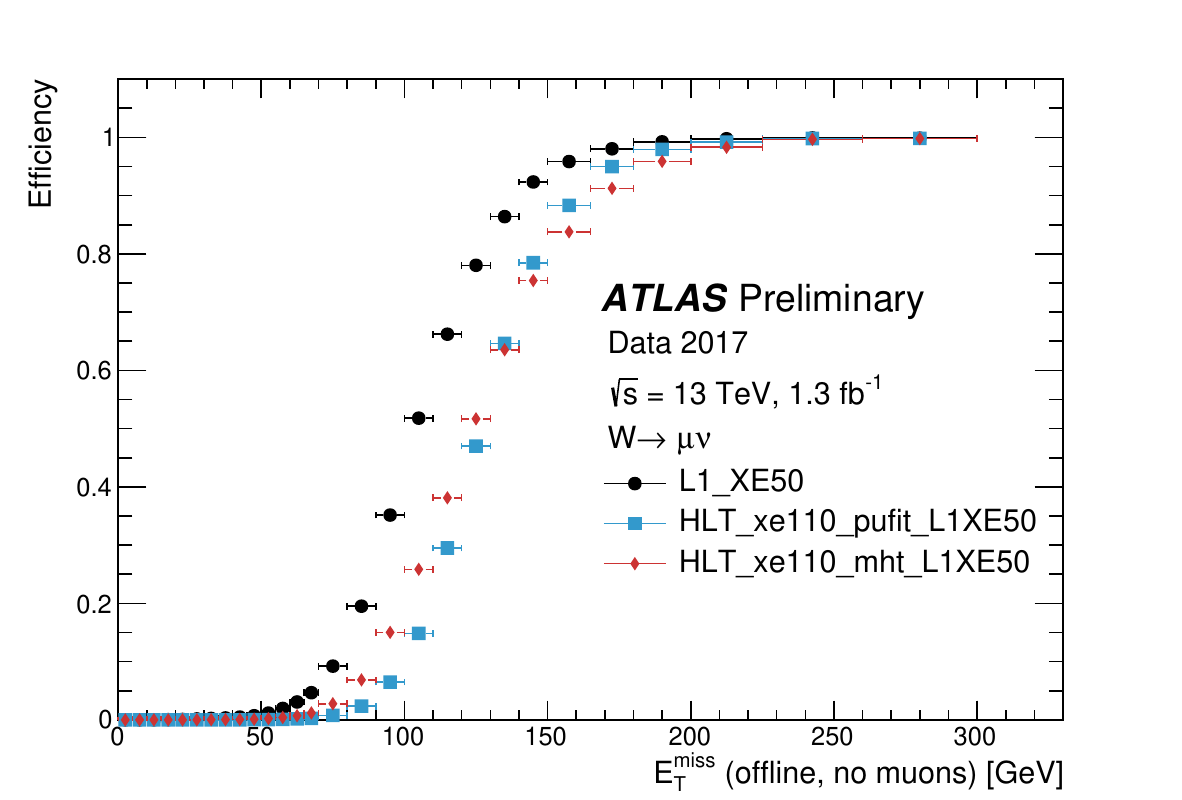}
\caption{The combined L1 and HLT efficiency of the missing transverse energy triggers HLT\_xe110\_pufit\_L1XE50 and HLT\_xe110\_mht\_L1XE50 as well as the efficiency of the corresponding L1 trigger (L1\_XE50) are shown 
as a function of the reconstructed $E_T^{miss}$ (modified to count muons as invisible)~\cite{bib:trigpub}. The events shown are taken from data with a $W\rightarrow\mu\nu$  selection to provide a sample enriched in real 
$E_T^{miss}$*.}
\label{fig:meteff}
\end{figure}

The largest challenges however probably come from the $E_T^{miss}$ triggers which require information about the entire detector, but which is also highly sensitive to pile-up. To benefit from the pile-up removal from jet
energy measurements at the HLT, an offline-like $E_T^{miss}$ was developed using trigger jets as input (MHT) rather than calorimeter cells or topoclusters. While such a reconstruction algorithm performed very well in 2015 and in 
early 2016, it rapidly became clear that this was not sufficient: the MHT algorithm is exponentially dependent on the pile-up increase. During the shutdown between Run-1 and Run-2 another algorithm was developed that was
suppressing pile-up energy on an event-by-event basis beyond what is reconstructed in the jets. This algorithm uses topoclusters energy in regions of the calorimeter where the hadronic activity is less intense,
and then performs a fit, under the assumption that the total pile-up does contribute to no net $E_T^{miss}$, to estimate the pile-up contribution to regions of the detector where the hadronic activity of the main process is likely to be 
situated~\cite{bib:atltriperf2015}. As can be seen in Fig.~\ref{fig:pufit}, this algorithm (PuFit) succeeded in linearizing the $E_T^{miss}$ rate dependence on pile-up, allowing much lower thresholds than would be otherwise possible. 
As can be seen in Fig.~\ref{fig:meteff}, the PuFit algorithm is even a little bit more efficient than the MHT algorithm, despite being much more different than the offline $E_T^{miss}$ reconstruction algorithm. Note that because of 
L1 improvements presented above, the L1 threshold is kept so low (50 GeV) that the only source of inefficiency with respect to offline comes from the much more precise HLT algorithms. 

\section{Conclusion}

Large statistic data samples constitute one of the key ingredients for exploring new physics and performing high-precision measurements. To do this, the LHC luminosity is continually increased. This 
constitutes a challenge for data-taking. To cope with the high luminosity and pile-up conditions of LHC in Run-2, the ATLAS TDAQ system went through a series of hardware, firmware, and software upgrades. 
At L1, these improvements led to a 30\% bandwidth increase as well as in the capacity to efficiently select events with a broader range of topologies. At the HLT, they allowed for the development of more 
performant algorithms deployed on all events such as the multistage track reconstruction and the full calorimeter scan topocluster formation with powerful pile-up mitigation. Information from tracks and calorimeter clusters are then 
used to efficiently select events with electrons, muons, taus, jets, and $E_T^{miss}$, often even exploiting multivariate techniques. Stronger correlations in the energy measurements of objects 
reconstructed online and offline have been observed compared to Run-1. Thanks to all these improvements, ATLAS succeeded in selecting with high performance the events needed for the success of its 
Run-2 physics program. New challenges are now awaiting for Run-3!
\\

\footnotesize{* From ATL-DAQ-PROC-2018-005. Published with permission by CERN.}






%

\end{document}